\documentstyle[pra,aps]{revtex}

\title{SOME REMARKS ABOUT INTRINSIC PARITY IN RYDER'S DERIVATION OF THE DIRAC
EQUATION\footnote{Published in American  Journal of Physics {\bf 63}, 177-178
(1995)}}

\author{Fabi\'an H. GAIOLI and Edgardo T. GARCIA ALVAREZ  \\
{\it Departamento de F\'{\i}sica, Facultad de Ciencias Exactas y Naturales, }
\\ {\it Universidad de Buenos Aires, 1428 Buenos Aires, Argentina} 
\\ {\it and }
\\ {\it Instituto de Astronom\'\i a y F\'\i sica del Espacio,} 
\\ {\it C.C. 67, Suc. 28, 1428 Buenos Aires, Argentina}}
\vspace{1.0in}

\date{\today}

\begin{document}

\maketitle

%\begin{abstract}

%\end{abstract}
\draft
\vspace{2.0in}
Pacs numbers: 03.65.Pm
\narrowtext
\twocolumn
\vspace{1.0in}

%\section{}

Ryder in one of  the  most  nowadays  widely  used textbooks on quantum field
theory \cite{ry}, has recently proposed  an  alternative  derivation  of  the
Dirac equation.  His proposal contains,  in  essence,  the  steps followed to
derive the plane wave solutions of the  free  Dirac  equation in an arbitrary
frame from the plane wave solutions in the  rest  frame,  but in the opposite
direction \cite{me}.  We consider that the underlying idea  in his derivation
has  a  significant  pedagogical  value,  but  it  omits important conceptual
features,  as  e.g., the physical meaning of the negative energies \cite{ry2}
and  the relative intrinsic parity of the elementary particles \cite{pi}.  In
this comment  we  repeat  Ryder's  derivation  amending such omissions, which
becomes a more  instructive  derivation since it reveals the physical content
of the Dirac equation.    Moreover,  we  explicitly  show  the  few  physical
hypotheses which are enough to uniquely determine it. 

Let us outline Ryder's argument which corresponds to $\epsilon =  1$  in  the
following  formulas.    We are searching for a wave equation for  spin  $1/2$
particles,  so  let us consider Weyl chiral spinors.  It is well  known  that
right  and    left  handed  Weyl  spinors  are  two  spin  $1/2$  irreducible
representations  of  the    Lorentz    group    interchanged    under  parity
transformation, which transform under boosts as \cite{do,bj} 

\begin{eqnarray}
\varphi_{R\atop                L}(\vec{v})&=&\exp[\pm\frac{1}{2}\vec{\sigma}.
\vec{n}\phi]\varphi_{R\atop L} (0) \nonumber
\\ 
&=&       [\cosh(\frac{\phi}{2})\pm      \vec{\sigma}.                \vec{n}
\sinh(\frac{\phi}{2})]\varphi_{R\atop L}(0), \label{rl}
\vspace{0.5in}
\end{eqnarray}
where  $\vec{\sigma}=(\sigma_1,\sigma_2,\sigma_3)$  are  the Pauli  matrices,
$\vec{v}=v\vec{n}$ is the velocity and $\phi$  the  additive parameter of the
boost ($\tanh \phi=v$;  \ $c=1$).  Supposing now that $\varphi_{R\atop L}(0)$
are  the  spinors  corresponding  to  the  frame  where  a  particle  (or  an
antiparticle)  is  at  rest.    Using  trigonometric algebra we  can  rewrite
(\ref{rl}) as

\begin{equation}
\varphi_{R\atop  L}(\epsilon E_p, \epsilon  \vec{p})=\Bigg[\sqrt{\frac{\gamma
+1}{2}}\pm    \vec{\sigma}.\epsilon       \frac{\vec{p}}{p}\sqrt{\frac{\gamma
-1}{2}}\Bigg]\varphi_{R\atop L}(\epsilon m, 0),\label{rl2} \vspace{0.5in}
\end{equation}
where         $\vec{p}=\epsilon        m\gamma\vec{v}$,                $m>0$,
$\gamma=(1-v^2)^{-\frac{1}{2}} = E_p/m$, and $\epsilon =+1$ for particles and
$\epsilon = -1$ for antiparticles.   

The  well  known  relativistic energy-momentum  relation  for  particles  and
antiparticles, 

\begin{equation}
E=\epsilon E_p=\epsilon \sqrt{p^2+m^2},\label{er}
\vspace{0.5in}
\end{equation}
where  $E$ represents the energy of  both  particles  and  antiparticles  [we
assume  that the antiparticles has negative energy;    see  the  comment  (a)
below], allows to put Eq.  (\ref{rl2}) in the form,  

\begin{equation}
\varphi_{R\atop    L}(\epsilon    E_p,    \epsilon    \vec{p})=\frac{E_p+m\pm
\epsilon\vec{\sigma}.  \vec{p}}{\sqrt{2m(E_p+m)}} \varphi_{R\atop L}(\epsilon
m, 0).\label{rl3}
\vspace{0.5in}
\end{equation}

Let us assume for a moment that, in the rest frame, 

\begin{equation}
\varphi_R(\epsilon m, 0)=\epsilon \varphi_L(\epsilon m,0).\label{pi}
\vspace{0.5in}
\end{equation}
(We will later discuss the  physical  meaning of this statement.) Then, using
the algebra of Pauli matrices we can immediately verify that 

\begin{equation}
\varphi_{R\atop  L}(\epsilon  E_p,  \epsilon  \vec{p})=\frac{\epsilon  E_p\pm
\vec{\sigma}.\vec{p}}{m}\varphi_{L\atop       R}(\epsilon    E_p,    \epsilon
\vec{p}).\label{pi2}
\vspace{0.5in}
\end{equation}

The system of equations (\ref{pi2}) is  nothing  else  than  another  way  of
writing the ``Dirac equation,'' as was originally proposed by Weyl \cite{we}, 

%\begin{mathletters}
%\label{allequations}
\begin{equation}
(\epsilon E_p +\vec{\sigma}.\vec{p})\varphi_L=m\varphi_R, \label{ea}
\end{equation}
\begin{equation}
(\epsilon E_p -\vec{\sigma}.\vec{p})\varphi_R=m\varphi_L. \label{ea2}
\vspace{0.5in}
\end{equation}
%\end{mathletters}
This  can  be  easily  verified,  defining  the  $\gamma$  matrices  and  the
four-spinor $\psi$ by

\begin{equation}
\gamma^0=\left(\matrix{0&1\cr
1&0\cr}\right), \ \ \ 
\vec{\gamma}=\left(\matrix{0&-\vec{\sigma}\cr
\vec{\sigma}&0\cr}\right), \ \ \ 
\psi={\varphi_R \choose \varphi_L}, \label{md}
\vspace{0.5in}
\end{equation}
and rewriting Eqs. (\ref{ea}) and (\ref{ea2}) in a more standard way,

\begin{equation}
(\gamma^{\mu}p_{\mu}-m)\psi(p_{\mu})=0,\label{ed}
\vspace{0.5in}
\end{equation}
where  $p_{\mu}=(E,-\vec{p})$  \cite{ry3}.    However, strictly speaking, Eq.
(\ref{ed}) is  not  the  Dirac  equation  yet, but it could be obtained, in a
heuristic  way,  replacing    $p_{\mu}$    by    the    ``operator''    $\hat
p_{\mu}=i\partial_{\mu}$ \ ($\hbar =1$) (see Ref.  \cite{afa}),

\begin{equation}
(i\gamma^{\mu}\partial_{\mu}-m)\psi_D(x^{\mu})=0,\label{ed2}
\vspace{0.5in}
\end{equation}
where   $\psi_D(x^{\mu})=\psi(p_{\mu})e^{-ip_{\mu}x^{\mu}}$.    Nevertheless,
this  result    cannot  be  rigorously  reached,  unless  one  considers  the
inhomogeneous Lorentz group from the very beginning. 

Let us now examine the hypotheses which lead to Eq. (\ref{ed}) in detail. 

(a) We have  assumed the standard formulas of classical relativistic dynamics
[e.g.,  Eq.    (\ref{er})],  associating  particles  and  antiparticles  with
positive  and  negative  energy values,  according  to  St\"uckelberg-Feynman
interpretation \cite{sf}. 

(b) We have also assumed that the desired equation must describe a spin $1/2$
system,  so  we have chosen spin $1/2$  representations  of  the  homogeneous
Lorentz group.  In this way we have derived the Weyl equations (\ref{ea}) and
(\ref{ea2}).    At  this point from these equations, making  the  replacement
$p_{\mu}$  by  $\hat  p_{\mu}$,  we  can  build a first order  equation  [Eq.
(\ref{ed2})] for  a  four-components  wave  function $\psi_D$ following Dirac
\cite{di}, which is  the  direct  sum  of  the  two Weyl representations [Eq.
(\ref{md})].  Alternatively, following  Feynman  and  Gell-Mann \cite{fg}, we
can combine Eqs.  (\ref{ea}) and (\ref{ea2}) to form a second order equation,
in which is necessary to consider  only one of the chiral two-components wave
functions.   

(c)  We  want  finally  to remark that  Eq.    (\ref{pi})  is  equivalent  to
postulating  that  the relative intrinsic parity \cite{pi} of  particles  and
antiparticles is opposite.  This fact means that, for  instance,  considering
that  $\varphi_{R\atop L}(\epsilon m, \epsilon \vec{p})$ represents the state
of the spin $1/2$ system, we can choose  

\begin{equation}
P\varphi_{R\atop L}(\epsilon  m,  0)=\epsilon  \varphi_{R\atop L}(\epsilon m,
0),\label{prr}
\vspace{0.5in}
\end{equation} 
where $P$ is the intrinsic parity operator. However, we know that 

\begin{equation}
P\varphi_{R\atop   L}(\epsilon  m,    0)=\varphi_{L\atop    R}(\epsilon    m,
0).\label{prl}
\vspace{0.5in}
\end{equation}
Then, comparing the two  sets of equations we obtain Eq.  (\ref{pi}).  (It is
easy to check that in each set of equations there is a redundant one.) 

To conclude, we have shown that  the  derivation exposed above highlights the
essential content of the Dirac equation [hypotheses (a), (b), and (c)].  This
fact contrasts with the standard derivation introduced in classical textbooks
\cite{bd},  which follows Dirac's original idea \cite{di} of linearizing  the
Klein-Gordon equation, i.e., to obtain a Schr\"odinger equation putting on an
equal  footing  both the temporal and the spatial derivatives by relativistic
invariance considerations \cite{afa}.    However,  as  it  was later admitted
\cite{ke}, the linearization condition  does  not uniquely lead to spin $1/2$
systems.  In Dirac's original  scheme,  the appearance of the spin $1/2$, the
negative energies (which are common to all spins), and the relative intrinsic
parity, i.e., conditions (a), (b), and (c),  came  as unexpected consequences
\cite{era}.  (This is what brought about the  broad  impact of Dirac's work.)
But now, a retrospective view, based on the experimental  facts, allows us to
uniquely  obtain the desired relativistic equation for a massive, spin  $1/2$
system.  Here mainly lies the already mentioned pedagogical value of  Ryder's
modified  derivation  of  the  Dirac  equation  that  we have revised in this
comment. 

\vspace{0.3in}
\noindent
\section*{Acknowledgments}

We  would  like  to  thank Professor Diego Harari for pointing out to us  his
doubts about hypothesis (\ref{pi}) in Ryder's original derivation.

\end{document}